\documentclass[acmsmall]{acmart}

\usepackage{booktabs} 
\usepackage{tabularx} 

\usepackage{titlesec}
\usepackage{float}
\titleformat{\paragraph}
{\normalfont\normalsize\bfseries}{\theparagraph}{1em}{}
\titlespacing*{\paragraph}
{0pt}{3.25ex plus 1ex minus .2ex}{1.5ex plus .2ex}

\usepackage{graphicx}
\usepackage{lscape}
\usepackage[figuresright]{rotating}
\usepackage{tabularx}
\usepackage{booktabs}

\newcolumntype{C}{>{\centering\arraybackslash}X}

\acmJournal{CSUR}
\acmVolume{xx}
\acmNumber{xx}
\acmArticle{xx}
\acmYear{2023}
\acmMonth{12}

\begin{document}
\title{Machine Learning Techniques for Sensor-based Human Activity Recognition with Data Heterogeneity - A Review}

\title{Machine Learning Techniques for Sensor-based Human Activity Recognition with Data Heterogeneity - A Review}

\author{Xiaozhou Ye}
\affiliation{%
  \institution{Department of Electrical, Computer, and Software Engineering, The University of Auckland}
  \city{Auckland}
  \country{New Zealand}
}

\author{Kouichi Sakurai}
\affiliation{%
  \institution{Department of Informatics, Kyushu University}
  \city{Fukuoka}
  \country{Japan}
}

\author{Nirmal Nair}
\affiliation{%
  \institution{Department of Electrical, Computer, and Software Engineering, The University of Auckland}
  \city{Auckland}
  \country{New Zealand}
}

\author{Kevin I-Kai Wang}
\affiliation{%
  \institution{Department of Electrical, Computer, and Software Engineering, The University of Auckland}
  \city{Auckland}
  \country{New Zealand}
}

\begin{abstract}
Sensor-based Human Activity Recognition (HAR) is crucial in ubiquitous computing, analysing behaviours through multi-dimensional observations. Despite research progress, HAR confronts challenges, particularly in data distribution assumptions. Most studies often assume uniform data distributions across datasets, contrasting with the varied nature of practical sensor data in human activities. Addressing data heterogeneity issues can improve performance, reduce computational costs, and aid in developing personalized, adaptive models with less annotated data. This review investigates how machine learning addresses data heterogeneity in HAR, by categorizing data heterogeneity types, applying corresponding suitable machine learning methods, summarizing available datasets, and discussing future challenges.
\end{abstract}


\maketitle

\section{Introduction}
Human activity recognition (HAR) is an important topic in ubiquitous computing to provide intelligent service and is designed to classify the activity of a human or group of humans. There are typically two types of HAR, namely, sensor-based and camera-based HAR. The wide penetration of embedded and wearable devices makes sensor-based HAR very popular and has been applied in a wide variety of applications, including ambient assisted living \cite{lago2019learning}, fitness and sports \cite{reiss2010activity}, rehabilitation \cite{conci2019utilizing}, security surveillance \cite{li2020fusing}, health monitoring \cite{van2010activity}, etc. Except for the sensor-based HAR, camera-based HAR is another HAR research direction that involves using cameras or video data to recognise human activities. While camera-based approaches may acquire richer information, there are also some obvious drawbacks such as low-light conditions, obstructed camera viewpoint and privacy concerns. Therefore, this review focuses on sensor-based HAR and its associated challenge in data heterogeneity.

Currently, most sensor-based HAR research assumes that the test and training samples satisfy the hypothesis of IID (Independent and Identically Distributed) \cite{chen2021deep} to ensure the generalization ability of the sensor-based HAR model. However, the practical situation is often different from the ideal scenario in the experimental environment, and the performance of these trained models will also decline dramatically. For example, different people have different physiological characteristics such as age, weight and height, and are highly likely to exhibit different activity data distributions. This phenomenon of different data distributions is called data heterogeneity \cite{chang2020systematic}. Data heterogeneity commonly occurs in embedded and IoT sensors, where collected datasets have non-uniform distribution. While being a common challenge to the AI community, the types of data heterogeneity are highly dependent on the applications \cite{wang2017heterogeneous}. The ability to cope with the data heterogeneity in sensor-based HAR can lead to the activity model performance improvement with less computation cost and help the construction of a personalized, adaptive and robust model with less annotation effort \cite{wang2018stratified}. Therefore, an increasing number of researchers are investigating how to handle sensor-based HAR data heterogeneity to reduce the time and effort required to adapt activity recognition algorithms to various users and scenarios, to improve the robustness and versatility of sensor-based activity recognition systems, and to reuse previously generated knowledge effectively.

In this review, the research questions we seek to answer are 1) what are the types of data heterogeneity in sensor-based HAR, and 2) what are the state-of-the-art machine learning techniques developed for addressing sensor-based HAR data heterogeneity. This review analyses the state-of-the-art research work and provides some insights into the suitable choice of machine learning approaches under the different data heterogeneity issues. Moreover, this paper discovers areas that still require further investigation. To the best of our knowledge, it is the first review that specifically focuses on the data heterogeneity issues for sensor-based HAR applications.

In order to keep a clear focus, we have applied some constraints to our survey. This review focuses on the sensor-based HAR data heterogeneity issues and camera-based HAR is not considered in this study unless it is used in conjunction with sensor-based approaches. Also, this review specifically focuses on sensor-based HAR for general physical and daily activities such as walking, jumping, taking shower, washing dishes, etc. Application specific activities such as fall detection or human disease/ill responses are not covered in this review. 

The remainder of the paper is structured as follows. Section 2 summarizes and categorizes the different types of data heterogeneity in sensor-based HAR, followed by some background introduction to the typical machine learning paradigms. Section 3 to 7 review different machine learning paradigms in different types of sensor-based HAR data heterogeneity, namely, \textbf{data modality heterogeneity}, \textbf{streaming data heterogeneity}, \textbf{subject data heterogeneity}, \textbf{spatial data heterogeneity} and \textbf{general framework for multiple heterogeneities}, respectively. Section 8 introduces the available public datasets for various types of data heterogeneity in sensor-based HAR. Finally, the future directions and conclusions are presented in Section 9.

\section{Data Heterogeneities and Machine Learning Paradigms in Sensor-based HAR}
In this section, we introduce the types of data heterogeneities, and the related machine learning paradigms used in this paper. To solve the issue of data heterogeneity in sensor-based HAR, several machine learning paradigms are applied by the research community. The focus of this review is on how these paradigms could contribute to solving the issue of sensor-based HAR data heterogeneity.

\subsection{Sensor-based HAR with Data Heterogeneities}
Sensor-based HAR aims to automatically identify and classify human activities using data collected from various sensors. HAR can provide valuable insights into how humans move, behave and interact with their environment. We follow the definition of sensor-based HAR as previous work \cite{bulling2014tutorial}\cite{abdallah2018activity}\cite{wang2019deep}. Here, we introduce the sensor-based activity categories and the sensor-based activity recognition data heterogeneity categories. We refer to the previous work of sensor-based HAR taxonomy \cite{chen2021deep}, and give the four categories of sensor-based activity.

\textbf{Atomic activity} \cite{li2020deep} refers to a basic or simple activity that is a component of a more complex activity. An atomic activity is a fundamental action that often lasts for a very short time. It is normally treated as the minimum component of human activity. For example, crossing arms forward, waist bends, kicking or knee bending can be considered atomic activity. \textbf{Daily living activity} refers to the repeated activities that people perform on a daily living. They often involve performing the same action in the same way over and over again. For example, brushing teeth involves the repeated movement of a toothbrush in a specific pattern. \textbf{Sports fitness activity} refers to the physical movements and exercises performed during sports and fitness activities. Repetitive activities are often a key component of sports and fitness training, as they help to build strength, endurance, and muscle memory. These activities may include running, cycling, weightlifting, and other exercises that are designed to improve physical fitness. \textbf{Composite activity} \cite{chen2021deep} can be defined as a sequence of sub-actions and have higher-level semantics. Composite activity recognition is a more challenging task than recognising simple activities because it requires not only detecting human body movements but also taking into account contextual information about the surrounding environment. For example, "making coffee" can be represented as a sequence of simple activities that happens in a kitchen environment. Table~\ref{tab_categories_of_sensor_activity} shows the four categories of sensor-based activity with the corresponding activities.

\begin{table}[h!] \scriptsize
\caption{Categories of sensor-based activity.\label{tab_categories_of_sensor_activity}}
\begin{tabular}{cc}
\begin{tabularx}{\textwidth}{CC}
\toprule 
Activity Category & Activities \\
\midrule 
Atomic activity & right arm throw, two hands front clap, frontal elevation of arms, waist bends forward, kicking, pocket out, flexion of the leg up, knees bending, draw zigzag, push, pull, slide, stand up, sit down, etc. \\ \hline
Daily living activity & brush teeth, talking on phone, cleaning, eating walking, sitting, standing, lying, etc. \\ \hline
Sports fitness activity & running, skiing, running, jogging, biking, playing basketball, exercising on stepper, rowing, rope jumping, dumbbell curl, squat upright row, Nordic walking, flick kicks, climbing, etc.  \\ \hline
Composite activity & make coffee, groom, cooking, clean counter tops, bathe, preparing breakfast, empty room, attend a presentation, housekeeping, using the toilet, personal hygiene, open close door, etc.\\ 
 \bottomrule
 \end{tabularx}
\end{tabular}
\end{table}

\begin{figure}[h!]
\includegraphics[width=0.8\textwidth]{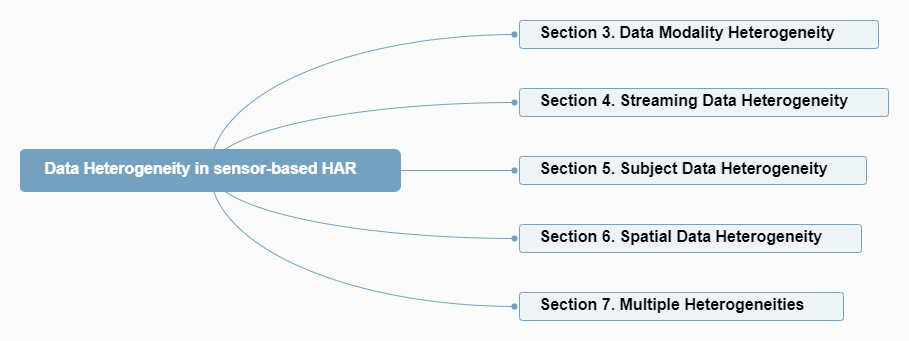}
\caption{Sensor-based HAR Data Heterogeneity Categories.\label{heterogeneity_category}}
\end{figure}

Data heterogeneity is a commonly occurring challenge in IoT data, where collected datasets have non-uniform distribution \cite{wang2017heterogeneous}. Considering sensor-based HAR applications, there are various sources of data heterogeneity: Firstly, data generated from a variety of sensor devices, which is called \textbf{data modality heterogeneity}. Different sensor types, platforms, manufacturers and modalities may cause the different data formats and distribution. Secondly, the dynamic streaming data pattern changes against time if the properties vary under a certain influence, which is called \textbf{streaming data heterogeneity}. For example, the walking pattern of the same person may affect by health and physical status. Thirdly, the behaviour differentiation between different people may also be significant, which is called \textbf{subject data heterogeneity}. For instance, some people may make coffee in the evening, while others may drink coffee in the morning. Fourthly,  different sensor networks in different body positions or different environmental layouts in smart homes may lead to different data distribution, which is called \textbf{spatial data heterogeneity}. Finally, the sensor-based HAR mixed data heterogeneity may happen that requires a general framework to handle the issue. Figure~\ref{heterogeneity_category} shows the five sensor-based HAR data heterogeneity categories.

\subsection{Machine Learning Paradigms} 

As for all the above mentioned different types of sensor-based HAR data heterogeneity issues, transfer learning is the most common machine learning paradigm that can be applied. \textbf{Transfer Learning} \cite{cook2013transfer} \cite{zhuang2020comprehensive} is a machine learning paradigm that captures knowledge from one predefined problem and applies it to a different problem. This paradigm covers a wide range of scenarios, and all transfer learning techniques emphasize the learning direction from the source domain(s) to a target domain and assume the knowledge in the source domain(s) could be helpful in the target domain. Transfer learning is particularly suitable for dealing with sensor-based HAR data heterogeneity problem and is the dominant method. In the sensor-based HAR field, typically, existing dataset(s) with labelled data can be seen as source domain(s). Then, a new dataset that the system has never seen before, such as a new activity, a new user, a new environment or a new device, is considered a target domain.

In the data modality heterogeneity field, multi-view learning is particularly suitable because each sensor modality can be considered as a distinct view to provide knowledge. \textbf{Multi-view Learning} \cite{xu2013survey} focuses on data representation via multiple distinct views to improve the generalization performance. This learning paradigm is based on the idea of two heads are better than one that integrates the advantages of different views. It aims to learn only a single task with all the views as opposed to multi-task learning which learns for many different tasks.

In the streaming data heterogeneity field, continual learning, zero-shot learning and few-shot Learning are the three common machine learning paradigms. \textbf{Continual Learning} \cite{de2021continual}, or lifelong learning, or incremental learning can process the flow of information over time and update the model continuously, and retain, integrate and optimize existing knowledge while absorbing new knowledge from emerging samples. This  paradigm focuses more on multiple sequential domains and tasks from a temporal dimension and implicitly copes with the different distributions and data heterogeneity across domains.

\textbf{Zero-shot Learning} \cite{wang2019survey} is another subfield of machine learning that focuses on classifying emerging classes. Typically, there are some labelled training samples and these data’s classes are called seen classes. However, several unlabelled testing samples of different classes are referred to as the unseen classes in the label space. This paradigm aims to identify these unseen classes. In zero-shot learning, the source domain feature space of training instances is the same as the target domain feature space of testing instances. However, the source task label space (the seen class set) is different from the target task label space (the unseen class set). Zero-shot learning is suitable for the scenario of new activities occurring.

\textbf{Few-shot Learning} \cite{wang2020generalizing} aims to train a model using only a few training examples based on prior knowledge of several similar tasks. One particular example of this is the ‘K-shot N-way’ classification, which means a trained classifier needs to classify N new classes using only K examples of each. If adding the restricted condition of only one instance for each class on few-shot learning, it becomes the so-called one-shot learning. In sensor-based HAR, scarce labelled data in the target domain is very common.

In the subject data heterogeneity field, federated learning, multi-task learning and ensemble learning are the three common machine learning paradigms. \textbf{Federated Learning} \cite{yang2019federated} is originally designed for distributed machine learning that emphasizes data security and privacy. Because of its good characteristic of sharing information cross all the users when doing the coordinated training, it attracts more and more attention for solving sensor-based HAR multiple users data heterogeneity issue. Normally, the data is heterogeneous across various users and this paradigm can be utilized to improve each user’s performance in a safe way. When the sensor-based HAR challenge is multi-user data heterogeneity, federated learning is worth a try.

\textbf{Multi-task Learning} \cite{zhang2021survey} pays attention to the collaborative and distributed learning of multiple tasks with shared models. The goal is to improve the performance of all the models with heterogeneous data. In contrast to transfer learning, they both aim to improve the performance of learners via knowledge transfer. However, transfer learning focuses on the one-way transfer of knowledge from the source domain(s) to the target domain(s). Multi-task learning focuses on knowledge transfer across tasks and it mainly takes advantage of the interconnections between tasks. Multi-task learning is especially useful in sensor-based HAR multiple users data heterogeneity scenario considering multiple users have multiple different tasks.

\textbf{Ensemble Learning} \cite{sagi2018ensemble} is to combine several base classifiers for improving a final model's generalization and robustness. Ensemble learning is based on the idea of achieving a better performance when there is significant diversity among the base classifiers. Each base classifier contains some knowledge, multiple base classifiers can provide more diverse and comprehensive knowledge. Therefore, the adjustment of the weight of different knowledge can be flexible to build a model for dealing with sensor-based HAR data heterogeneity issues.

In the spatial data heterogeneity field, \textbf{multi-view learning} can also be applied because it can handle environment layout heterogeneity synchronously across multiple houses. Moreover, domain generalization can be applied to handle sensor-based HAR multiple heterogeneity scenario, \textbf{domain generalization} \cite{wang2022generalizing} is a particular type of transfer learning that aims to learn a model that can generalize to a target domain with one or several different but related source domain(s). Compared to the common transfer learning training in which both the source and target domain data can be accessed, domain generalization can only access several source domain data for training, and target domain data is not accessible. Therefore, domain generalization aims to tackle more challenging and practical scenarios.

\begin{figure}[h!]
\includegraphics[width=\textwidth]{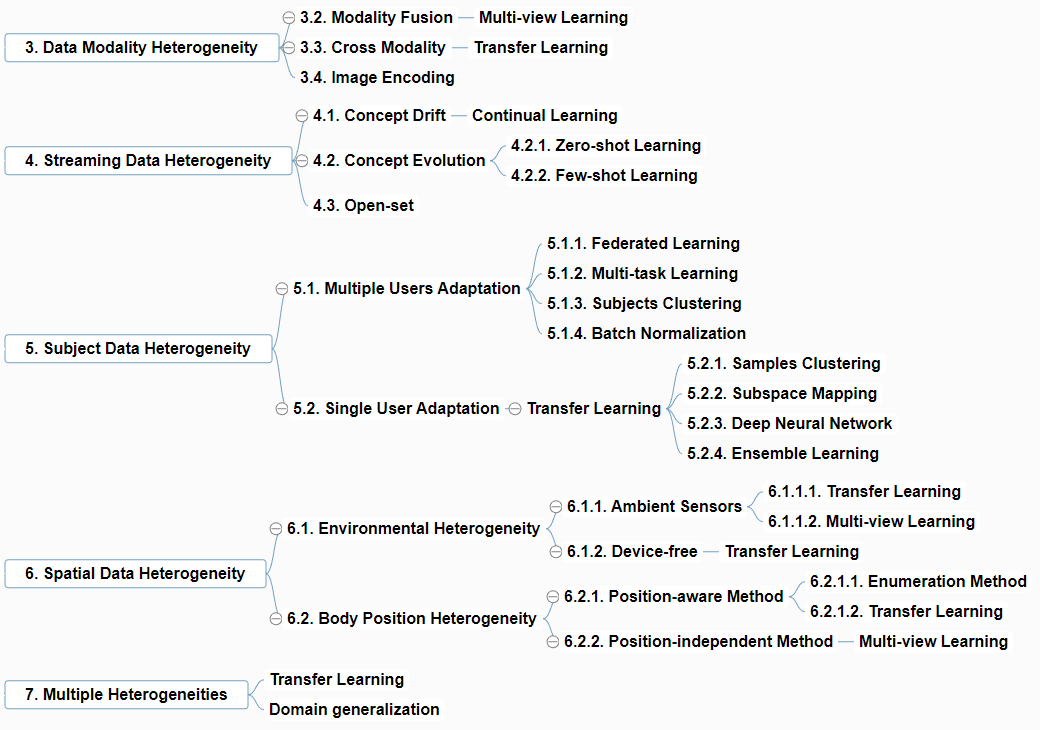}
\caption{Overall Machine Learning Techniques Framework.\label{overall_structure}}
\end{figure}

In the following sections, we will discuss different machine learning techniques applied in the five types of HAR data heterogeneities, which are summarized and illustrated in Figure~\ref{overall_structure}.

\section{Data Modality Heterogeneity}
Various types of sensors are utilized to implement sensor-based HAR in different scenarios as shown in Table ~\ref{tab2}. The performance of a system highly depends on these sensors as the source of the data. Machine learning techniques for addressing data modality heterogeneity can be mainly partitioned into modality fusion and cross modality. Modality fusion aims to combine and integrate multiple sensor modalities into a uniform form as multi-view learning. For instance, inertial measurement unit (IMU) sensor-based HAR can be combined with camera-based HAR to provide more board views for making up for insufficient IMU sensor-based HAR. Cross modality focuses on transforming from one modality to another modality to achieve certain goals under the learning paradigms of transfer learning. For example, camera-based HAR transfers its knowledge to sensor-based HAR for better performance of sensor-based HAR.

\begin{table}[h!] \scriptsize
\caption{Categories of Device Types and Sensor Modality.\label{tab2}}
\begin{tabularx}{\textwidth}{CCC}
\toprule 
 Sensor Type & Sensor Name & Activity Categories\\
\midrule 
Inertial Measurement Unit (IMU)  Sensors & Accelerometer, Gyroscope, Magnetometer & Atomic activity, Daily living activity, Sports fitness activity, Composite activity\\
\midrule
Ambient Sensors  & Pressure/Force, Infrared, Magnetic Switches/Contact, Ultrasonic, Light, Temperature, Humidity & Daily living activity, Composite activity\\
\midrule
Device-free Sensors          & RFID, Wi-Fi, Radar & Atomic activity,  Daily living activity  \\ 
 \bottomrule
 \end{tabularx}
\end{table}

\subsection{Sensor Modality}

We first introduce different categories of sensors. Sensor modality is classified into three types: inertial measurement unit sensors, ambient sensors, and device-free sensors. The category of sensor modality refers to the reviews \cite{lentzas2020non} \cite{wang2018device}\cite{liu2023sensor}. 

\textbf{IMU sensors} are the specific type of sensors that measures angular rate, force and sometimes magnetic field. They are portable and can be deployed in different body positions such as waist, ankle, arm, leg, back and chest to capture body context data directly. Inertial measurement unit sensors can also be embedded in smartphones that are carried in the pocket or held in hands to collect data. Inertial measurement unit sensors are very common in HAR research that have broad application scenarios such as daily activity recognition \cite{ohashi2018attributes}, sports activity recognition \cite{brunner2019swimming} and working activity recognition \cite{grzeszick2017deep}.

\textbf{Ambient sensors} are normally deployed in the environment at fixed positions to capture the interaction information between humans and the surroundings. In the HAR scenario, they are the essential devices in the smart home (or related) applications and can be embedded with extra position information. 

\textbf{Device-free sensors} wireless sensing research is the new trend in HAR. It shows that wireless signals can not only be utilized as a communication tool to exchange data but also has the capability to reflect the differences in various human behaviours and sense the variation of the surrounding environment. This type of sensor normally has a transmitter and receiver to send and receive electromagnetic waves. Different human activities lead to different reflected signal patterns. Device-free sensors can normally be deployed on the wall or ceiling of a room.

\subsection{Modality Fusion}

The purpose of modality fusion is to find a way to integrate various sensor modalities for serving the improvement of the sensor-based HAR system. The mainstream learning paradigm in modality fusion is multi-view learning due to its inherent characteristics that treats each sensor modality as a separate view. Then the multiple views synchronously observe the same activity and unite to finish the common activity classification task. There are mainly two types of modality fusion: Data-level/Feature-level fusion and classifier-level fusion. The human activities researched in modality fusion include stairs up, stairs down, walking, biking, standing, pressing buttons, plugging cables, rotating the chair, using hammer, clap, arm cross, bowling, tennis serve, baseball swing, etc.

\textbf{Data-level/Feature-level fusion} extracts features from each sensor and combines them to train a single classification model. The most common approach in feature-level fusion is aggregation which concatenates the extracted features or certain raw data from all sensors. Ehatisham et al. \cite{ehatisham2019robust} focuses on feature-level data fusion from two modalities: RGB/depth video camera and inertial body sensors. It extracted features separately and then concatenated the two feature vectors. These features contain densely extracted histograms of oriented gradient features from the camera and statistical time-series characteristics from wearable sensor data. Li et al. \cite{li2020fusing} combined the Kinect depth camera and wearable sensor as a complementary fusion to improve the performance. The depth data features and wearable sensor data are joined in a data-level fusion. Data contraction is realized by feature selection with metaheuristics methods. Finally, incremental learning is applied via the Hoeffding tree and the Swarm Decision Table. 

Inspired by the design of non-local block in self-attention deep networks, which is used to compute relations between different space-time locations of the same instance. Byvshev et al. \cite{byvshev2020heterogeneous} adapted the non-local block for heterogeneous data fusion of video and smart-glove. There are two main change points: (1) the changes from self-attention to cross-modal attention via two multi-modal inputs replacing the input in the original design, (2) adjusting the non-local operation in a way without the requirement of spatio-temporal dimensions match for data fusion. 

Compared to most of the work in this field with relatively big differences in the modalities such as ambient sensor and video, Stisen et al. \cite{stisen2015smart} did systematic research on the heterogeneity of 36 smartphones, smartwatches and tablets with various different device models from four manufacturers. The data heterogeneity mainly originates from sensor biases (poor sensor quality or accidental device damage), sampling rate heterogeneity (different device factory settings) and sampling rate instability (varying I/O load affect actual sampling rate). For mitigating the effect caused by heterogeneities, the kNN clustering method is applied to group similar devices and then a classifier is trained based on the devices’ sensor data in each cluster.

\textbf{Classifier-level fusion} aims to combine the classifiers which trained on each sensor modality to build a more robust model. This fusion strategy overcomes the feature-level fusion’s drawback of feature compatibility issues regarding heterogeneous sampling frequencies and configuration parameters. Garcia-Ceja et al. \cite{garcia2018multi} proposed a staking ensemble based multi-view learning method for modality fusion. It is different from traditional multi-view learning via aggregation which mixes the feature space from multiple views. This method builds individual models for each view and combines them using a stacked generalization. The highlight of this paper is the comparison of audio view, accelerometer view, aggregation view and multi-view stacking and the results show that multi-view stacking achieves the best performance. 

With the same stacking ensemble technique in \cite{garcia2018multi}, Chung et al. \cite{chung2019sensor} extended the meta-learner to the voting strategy and compare it with the stacking method. Besides, they also compared three machine learning models, i.e., Random Forest (RF), k-Nearest Neighbors (kNN) and Support Vector Machines (SVM) in stacking ensemble. Xue et al. \cite{xue2019deepfusion} proposed a deep learning architecture for a unified multiple modality fusion, which considered the different sensors’ data quality for information weighted adjustment and the correlations among various sensors. Sensor-representation module is a Convolutional Neural Network (CNN) structure to extract low dimensional features of the same size even with non-uniform heterogeneous input. Weighted-combination module is inspired by the idea of an attention mechanism, which estimates the quality of each sensor data and combines the information in a weighted manner. Cross-sensor module captures the correlation of the various sensor information to learn a more robust model via averaging operation.

\begin{table}[h!] \scriptsize
\caption{Modality Fusion Methods Comparison.\label{modality_fusion_compare_table}}
\begin{tabularx}{\textwidth}{CCCCC}
\toprule 
\textbf{Fusion Strategy} & \textbf{Techniques} & \textbf{Pros} & \textbf{Cons}  & \textbf{Activity Categories}\\
\midrule
Data-level/Feature-level Fusion & concatenation\cite{ehatisham2019robust}\cite{li2020fusing} 
self-attention\cite{byvshev2020heterogeneous} clustering\cite{stisen2015smart} & consider the weighted contributions of all the different modality features/data & need to unify different modality features/data into the same format & atomic activity, daily living activity, sports fitness activity \\ 
\midrule
Classifier-level Fusion & stacking ensemble\cite{garcia2018multi} voting strategy\cite{chung2019sensor} attention weight\cite{xue2019deepfusion} & decouple the features in each modality make model training convenient & deal with inconsistent results across classifiers & daily living activity \\ 
\bottomrule
\end{tabularx}
\end{table}

As Table ~\ref{modality_fusion_compare_table} shows, for the modality fusion approaches, Data-level/Feature-level fusion follows a fine-grained fusion way, it considers all the features/data in different modalities together. After the learning process, the weighted contributions of the features/data for the final task can be achieved. In this way, only one classifier is required, which saves training time and computation resources. However, unifying different modality features/data into the same format is an extra step. The different modality features/data compatibility issue needs to be solved carefully. 
Classifier-level fusion approach is a coarse-grained fusion way that focuses on the multiple classifiers' co-decision. It is convenient that there is no need to consider the different modality compatibility issue and each modality can be trained based on its own common method. However, the different classifiers may have their bias and how to deal with inconsistent results across classifiers is another problem. For the activity categories, data-level/feature-level fusion focuses on atomic activity, daily living activity and sports fitness activity, while classifier-level fusion considers the daily living activity.

\subsection{Cross Modality}	
Cross modality aims to transfer knowledge from one modality to another modality to assist the goal attainment of different data modality. The main learning paradigm in sensor-based HAR cross modality is transfer learning because of the directional knowledge transformation, and the goal is transferring useful information from the source modality to the target modality. Cross modality knowledge transformation aims to build a bridge and find the latent relationship between different data modality for taking advantage of knowledge from the source domain. There are mainly four branches in sensor-based HAR cross modality approaches: deep neural network fine-tuning, multi-view transfer learning, knowledge distillation and generative network method.

Xing et al. \cite{xing2018enabling} proposed the \textbf{deep learning fine-tuning} method, which can transfer knowledge across vision, audio and inertial measurement unit (IMU). For the knowledge transfer, source domain data is fed into a pre-trained model for calculating the activation value in an intermediate layer as a cross-domain shared feature vector. Then target domain data is used to train a network to map the shared feature vector to a unified feature space across multiple domains. For the task transfer, if the tasks are the same between the source and target domain, the source domain model’s higher layers for classification can be re-used directly in the target domain model. If the tasks are different, the target model’s low and intermediate layers are frozen, and the target model’s higher layers will be retrained via limited target domain data. 

Feuz and Cook \cite{feuz2017collegial} introduced a \textbf{multi-view transfer learning} algorithm with or without labelled data in the target domain. Ambient sensors, smartphone/wearable sensors and video cameras use transfer learning to act as colleagues. For the availability of a small amount of labelled data in each view scenario, a separate weak classifier is trained for each view and then all classifiers select the unlabelled samples to add to the labelled set by a confidence score. For the unlabelled target domain data scenario, subspace learning is utilized to project the source domain and target domain onto a latent feature space via Principal Component Analysis (PCA) dimensionality reduction. After the subspace aligns between views via Procrustes analysis, projected data from the source view is used to train a classifier that tests projected data from the target view for data annotation. Finally, a classifier is trained via labelled data in the target view.

Kong et al. \cite{kong2019mmact} presented a \textbf{knowledge distillation} method for transferring knowledge from wearable sensor modality to video camera modality. Knowledge distillation allows the student network to capture the information provided by the ground truth labels and the finer structure learned by the teacher network. In the first step of generating the teacher model, multi-view learning is applied to multiple wearable sensors to train a classifier for each sensor, and then a weighted adaptive method is used to combine the classifiers according to their feature representation. In the second step, the teacher model’s knowledge is transferred to the student model via training classification loss and distillation loss. 

Zhang et al. \cite{zhang2020deep} worked on \textbf{deep generative networks} for transferring video data to synthetic accelerometer data. They proposed two algorithms: conditional generative adversarial network (cGAN) and conditional variational autoencoder-conditional GAN (cVAE-cGAN). For cGAN, a video encoder is used to compress the video clip into a feature vector. Then GAN models the conditional distribution of sensor data based on the video feature vector. cVAE-cGAN is an extended version of cGAN. A cVAE uses the information from the video to learn a conditional distribution of the sensor data based on the idea that prior knowledge from cVAE will improve the generative capability of GAN.

\begin{table}[h!] \scriptsize
\caption{Cross Modality Methods Comparison.\label{cross_modality_compare_table}}
\begin{tabularx}{\textwidth}{CCCCCC}
\toprule 
\textbf{Methods} & \textbf{Data Modality} & \textbf{Labelled Data Requirement} & \textbf{Pros} & \textbf{Cons} & \textbf{Activity Categories} \\ \hline
Fine-tuning\cite{xing2018enabling} & video camera, audio and wearable sensor & source domain + target domain & more efficient than training a model from scratch & fine-tuning may not be effective if a new task is very different from the original one & daily living activity, sports fitness activity\\ \hline
Multi-view Transfer Learning\cite{feuz2017collegial} &  Ambient sensor, smartphone/wearable sensor and video camera & source domain (+ target domain) & can be applied to the scenario without labelled target domain data & more views make the training more computationally expensive & composite activity\\ \hline
Knowledge Distillation\cite{kong2019mmact} & wearable sensor, video camera & source domain + target domain & the distilled model may generalize better to new data & may be loss of information & daily living activity, atomic activity\\ \hline
Generative Network\cite{zhang2020deep} & video camera, accelerometer & source domain + target domain & can produce a large amount of synthetic data & the generated data's quality may be lower than the real data & daily living activity, sports fitness activity\\
\bottomrule
\end{tabularx}
\end{table}

As Table ~\ref{cross_modality_compare_table} shows, for the modality fusion approaches, multi-view transfer learning is the only method that can be applied to the scenario of no labelled target domain data. It considers various classifiers' results and provides comprehensive views for the target domain's generation of pseudo labels to draw on the wisdom of the masses. However, more training resources are the drawback. In contrast, fine-tuning is a more computation resource-friendly method and it does not train the model from scratch. The drawback of fine-tuning is that the performance may drop dramatically if the new task is very different from the original task. Knowledge distillation trains the distilled model on the combination of the original model's predictions and the ground-truth labels, which may be able to improve the generalization to new data than the original model. However, this type of model compression technique may not be able to capture all of the information and nuances present in the original model. This could result in lower performance on certain tasks. The generative network can produce a large amount of synthetic target domain modality data, which can be useful for data augmentation. The generative network may not capture all of the nuances and details present in the real data, which could result in lower quality or unrealistic outputs. The corresponding human activity categories in each approach are also listed in Table ~\ref{cross_modality_compare_table}. For the activity categories, fine-tuning and generative network focus on daily living activity and sports fitness activity. The knowledge distillation approach aims to handle the daily living activity and atomic activity, while multi-view transfer learning focuses only on composite activity.

\section{Streaming Data Heterogeneity}

Sensor-based HAR often requires continuous data streaming. Typically, sensor data is a continuous and infinite data stream with a high sample rate and variable data distribution. However, many researchers assume the data distribution remains the same over time to simplify the problem. Dynamic changes in activities over time are an inherent and natural characteristic of the sensor data stream. Variation in existing activities or the emergence of new activities occurs in evolving activity data stream \cite{abdallah2018activity}. As for the example of variation in existing activities, people's activities in the morning may have different data distribution from the same activities at night because of muscular fatigue. As for the instance of the emergence of new activities, people may perform new activities such as learning new outdoor activities. There are mainly three categories of streaming data heterogeneity, namely concept drift, concept evolution, and open-set issue. 

\subsection{Concept Drift}

Concept drift \cite{gama2014survey} is a data heterogeneity issue in the temporal dimension, which refers to the distribution change over time. In the sensor-based HAR concept drift scenario, the historical data is typically considered the source domain and the emerging data is considered the target domain. Continual learning is the mainstream learning paradigm in concept drift because it is typically applied to unlimited streaming data. Various methods have been proposed that focus on identifying when the concept drift occurs and how to adjust the model to adapt to concept drift.

Roggen et al. \cite{roggen2013adarc} presented an adaptive continual learning framework. In comparison with typical supervised HAR models, this method has three additional steps for handling concept drift: self-monitoring, adaptation strategies and exploitation of available feedback. Self-monitoring discovers relevant changes in the system’s performance for identifying the possible concept drift point. Adaptation strategies adjust the parameters of the activity recognition model for acceptable performance after the concept drift happens. Available feedback includes the user, the activity-aware application, and external systems for guiding the model adaptation. Because of the general framework design, various machine learning techniques can replace the components in the adaptive activity recognition chain framework.

Abdallah et al. \cite{abdallah2015adaptive} introduced a two-phase method (i.e. online and offline) to handle data streams. In the offline phase, a cluster dictionary is generated that includes a set of clusters with their corresponding sub-clusters. Sub-clusters are the different patterns for each cluster. Then, an ensemble classifier is generated based on several hybrid similarity measures approaches for prediction. This classifier deploys an ensemble of four measures to assess the similarity of new data with the learned cluster dictionary. The online phase involves activity recognition and adaptation components. Upon the arrival of the new data stream, if each measure chooses a different cluster, it is highly possible that the concept drift has happened, and the data is required to be annotated. Then, the adaptation component updates the sub-cluster inside the corresponding cluster based on the newly labelled data to adjust the similarity distance of the four measures for solving concept drift.

Lima et al. \cite{lima2021nohar} presented a continual learning classification algorithm based on symbolic data generated by a discretization process using algorithms of SAX \cite{lin2007experiencing}. First, each chunk is a time window that goes through a discretization process, and the data is transformed into symbols. Then, these symbols in each chunk are represented via histograms based on their frequency distributions. This generated dictionary of frequency distributions is the feature of the corresponding activity class. In the activity recognition step, the algorithm compares the frequency distributions between the histograms of new data and the dictionary of histograms in each class. If there are changes in the frequency distribution of histograms caused, concept drift is considered to happen. In the adaptation step for handling concept drift, all existing histograms are retrieved to add some novel histograms that never appeared before, delete some histograms that do not occur any more and keep the histograms that are still effective after concept drift.

Meng et al. \cite{meng2017towards} proposed a knowledge-based continual learning method, which considered activity recognition and abnormal behaviour detection together. Here, abnormal behaviour is viewed as concept drift that has a different data distribution compared to normal behaviour. The unique point of this paper is the design of a dynamic daily habit component that aims to learn the habit of the elderly from their daily activities via the construction of two-layer tree architecture. The nodes in the first layer refer to the activity classes, and the probabilities of the user performing each activity in different time periods are calculated in the second layer. By doing so, the dynamic daily habit component can not only find the statistic pattern of the users’ daily habits but also serve as a knowledge base assisting the concept drift detection for users’ daily lives. The purpose of this work is to identify the concept drift, therefore there is no model adaptation step for solving concept drift.

\begin{table}[h!] \scriptsize
\caption{Concept Drift Methods Comparison.\label{concept_drift_compare_table}}
\begin{tabularx}{\textwidth}{CCCCCC}
\toprule 
\textbf{Methods} & \textbf{Labelled New Data} & \textbf{Feedback Level} &\textbf{Pros} & \textbf{Cons} & \textbf{Activity Categories}\\ \hline
Adaptive Activity Recognition framework\cite{roggen2013adarc} & Yes & high & more compatibility and scalability & need more human and outside guidance & sports fitness activity\\ \hline
Cluster + Ensemble Classifier\cite{abdallah2015adaptive} & Yes & median & rectify model effectively & need some human feedback & daily living activity, sports fitness activity\\ \hline
Data Symbolic Discretization + Frequency Distributions\cite{lima2021nohar} & No & low & less computation resource & sensor value compression uncertainty & daily living activity\\ \hline
Knowledge-based\cite{meng2017towards} & No & low & not require training data & may not flexible enough and lead to false detection & composite activity\\
\bottomrule
\end{tabularx}
\end{table}

As Table ~\ref{concept_drift_compare_table} shows, the adaptive activity recognition framework and cluster ensemble classifier method need certain feedback and labelled new data when concept drift happens. In this way, it adds some extra manual effort but can rectify the model effectively to achieve a guaranteed performance. The cluster ensemble classifier method includes some high confidence data based on its cluster similarity mechanism that reduces the degree of human feedback compared to the adaptive activity recognition framework. Adaptive activity recognition framework can be integrated into many machine learning techniques and components, which has better compatibility and scalability. The cluster ensemble classifier method and data symbolic discretization frequency distributions method essentially aim to find good feature representations that can be used on activity classification and different time period similarity measures. The data symbolic discretization frequency distributions method applies time series continuous data compression techniques that reduce the algorithm computation requirement dramatically. However, it also leads to loss of information, where a sensor value may be grouped into a different range of symbolic discretization that causes uncertainty issues affecting the following concept drift detection and adaptation. The generation of knowledge rules depends on expert experience and does not require any training data at all. However, a small data distribution variation may violate the rules and false detection may happen. The corresponding human activity categories in each approach are also listed in Table ~\ref{concept_drift_compare_table}. The adaptive activity recognition framework aims to solve sports fitness activity, while the cluster + ensemble
classifier has the extra category of daily living activity. The data symbolic discretization + frequency distributions method focuses on daily living activity, while the knowledge-based approach is designed for composite activity.

\subsection{Concept Evolution}

Concept evolution \cite{masud2010addressing} means that new activities appear in the continuous data stream, and the new classes need to be incorporated into the existing classes. Typically, people may perform new behaviours that never occurred in the past. Moreover, some rarely happened activities, or accidental events may happen unexpectedly, such as falling. Concept evolution aims to identify the concrete new activity instead of filtering and ignoring these new classes. Some methods in this area also belong to continual learning because of the characteristic of infinite streaming data. Some methods focus on zero-shot learning without labelled new data. Some methods relax the strict limits of zero-shot learning and focus on few-shot learning and one-shot learning with limited labelled new data.

\subsubsection{Zero-shot Learning}
Zero-shot learning applied to concept evolution focuses on capturing the semantic relations between old concepts and new concepts without labelled new concepts. Wang et al. \cite{wang2017zero} proposed a method for zero-shot learning via learning nonlinear compatibility function between feature space instances and semantic space prototypes. Here, prototype means a 0-1 vector; if instances of a class have an attribute, the corresponding digit in the class prototype is “1”. In the training phase, compatibility-based methods aim to learn a function measuring the compatible degree between a feature space instance and a semantic space prototype for building the similarity relationship of the two spaces. In the testing phase, the testing instance’s compatibility score is calculated with all unseen class prototypes and the unseen class with the highest score is assigned to the instance. 

 Machot et al. \cite{al2020zero} seek off-site assistance and considered the semantic distance between different activity words in the natural language semantic view. In the training phase, the names of the specific activities are encoded to the word vectors via Google Word2Vec representation. These word vectors are now the training labels, which is different from the common classification labels encoding way. Then, the sensor readings of the activities and the corresponding word vectors label are fed into a shallow neural network for model training. When concept evolution occurs, the new unseen sensor readings are sent to the neural network to get the output of a word vector. The word vector is applied to the nearest neighbour matching algorithm to find the most similar concept. This method takes advantage of the natural language semantic relationship to provide extra information for zero-shot learning. 

Different from the common methods of taking advantage of the semantic relationship between old and new concepts, \cite{hu2018novel} focuses on the assumption when new class data’s label is independent of the past labels. Some common time series features are utilized in measuring the relation between old concepts and new concepts. The Axis-Aligned minimum Bounding Box (AABB) is introduced into the random forest (RF). A separating axis theorem based splitting strategy is proposed with AABB, which enables the decision tree to be inserted with new nodes or split a leaf node without changing the original structure of the decision tree. From the geometry view, AABB is the box with the smallest surrounding space in which all the discrete points lie. To enable the insertion of a parent node, AABB is introduced to construct an incremental learning decision tree. It is used for finding the intersection between an existing node and new data. Moreover, the Separating Axis Theorem is introduced to find an appropriate attribute and position to split the decision tree when concept evolution happens. 

Hartmann et al. \cite{hartmann2022interactive} proposed an interactive method that asks the user to annotate the new data when new activities emerge. Then, the classification model is re-trained on-device with
new activities and the new model is loaded when a user decides to switch to the new recognition model.

\subsubsection{Few-shot Learning}

Compared to zero-shot learning, few-shot learning has loosened the restriction of no labelled data at all. With the limited labelled data, the model is able to adjust further to fit the new activity classes, and the performance of the system can be improved. Wijekoon et al. \cite{wijekoon2020knowledge} used a neural network with an attention mechanism to integrate new concepts into the existing model when concept evolution happened. During the training step, the data is divided into two parts: a small number of labelled data called support set and the rest of the labelled data called query instances. Then, a neural network is used to learn the common representative features for transforming both parts of the data into feature vectors. The cosine similarity metric is calculated for measuring the similarity between all query instances and support set instance pairs. In addition, an attention mechanism is utilized to estimate the class distribution. The pair mapping distance between the query instance and support set instances that belong to the same activity class is minimized via the neural network training. When concept evolution occurs, the new class data's distance to the support set will be large. Then, cosine similarity will be calculated between the new data to the support set and attention weight will be updated. 

The technique presented in \cite{ye2020evolving} integrated two advanced continual learning methods called Net2Net and Gradient Episodic Memory (GEM) to identify new concepts. When concept evolution occurs, Net2Net expands an already trained network by adding more neurons and layers to improve the learning ability to distinguish new concepts. The model is often extended from the last layer to fit new classes and from the second last layer to improve the learning capacity to discriminate a wider range of classes. GEM reduces the forgetting effect by balancing the performance of old and new classes and preventing disastrous forgetting of old concepts by controlling gradient updates. GEM ensures that the loss at previous tasks does not escalate after each parameter adjustment to avoid forgetting. 

Ye et al. \cite{ye2021continual} proposed a GAN based method that preserved old concepts without storage of historical data. When new concepts occur, samples generated from GAN combined with new data are used for classification. First, the GAN and a classifier which is a deep neural network are trained based on the current training data. When the emerging data contain new concepts, the GAN will generate samples for the previous activities with the learned latent structure. With the generated samples and the emerging data, the classifier is updated to recognise activities for both old and new concepts. The GAN will also be updated to be able to generate samples for the new activities. The above process will repeat and iteratively implement during the concept evolution process.

\subsubsection{Methods Comparison}

\begin{table}[h!] \scriptsize
\caption{Concept Evolution Learning Paradigms Comparison.\label{concept_evolution_compare_table}}
\begin{tabularx}{\textwidth}{CCCCCC}
\toprule 
\textbf{Learning Paradigms} & \textbf{Labelled New Data} & \textbf{Pros} & \textbf{Cons} & \textbf{Activity Categories}\\ \hline
Zero-shot Learning & No & no need for labelled new concepts & performance may drop dramatically due to weak relation between old concepts and new concepts & daily living activity, sports fitness activity, atomic activity, composite activity\\ \hline
Few-shot Learning & Yes & has certain guidance to learn new concepts & incomplete data distribution & daily living activity, sports fitness activity, atomic activity\\ 
\bottomrule
\end{tabularx}
\end{table}

As shown in Table~\ref{concept_evolution_compare_table}, zero-shot learning aims to learn new concepts without labelled data. The mainstream methods focus on mining the relation between new concepts and old concepts that utilizes historical knowledge for emerging knowledge. However, if the new concepts have no or weak relation between old concepts and new concepts, then the historical knowledge becomes less effective for the new concepts learning. Few-shot learning requires a certain amount of labelled data for a new concept. The limited amount of labelled data guides the learning direction. However, if the labelled new data are not representative enough and only cover a subset of the complete concepts, then the concept evolution direction may be misunderstood and the new concepts may not be learned properly. The corresponding human activity categories in each learning approach are also listed in Table ~\ref{concept_evolution_compare_table}. Both zero-shot learning and few-shot learning focus on daily living activity, sports fitness activity, and atomic activity. However, only zero-shot learning has been applied to the composite activity.

\subsection{Open-set}

Open set recognition \cite{geng2020recent} is a new and trending research direction. It considers the practical scenario of incomplete knowledge of the world that exists at training time, and unseen classes can be fed into the algorithm at testing time. The requirements of this type of approach are to accurately classify the seen classes and effectively reject unseen classes. Compared to concept evolution, open set recognition does not need to identify specific unseen classes and only needs to treat the unseen classes as a whole part. Based on the idea of constructing a negative set to represent the unseen activity space \cite{yang2019open}, and the open-set problem is simplified as “the seen classes plus one negative set class” classification problem. Moreover, a GAN deep learning model is proposed to learn to generate the negative set of synthetic data. After the training process, the discriminator part of GAN is the classifier for handling the open-set issue.

\section{Subject Data Heterogeneity}

Another commonly occurring sensor-based HAR data heterogeneity is caused by different subjects. For example, different people have different walking patterns considering the variations of age, weight and height. The relevant techniques can be considered mainly in two categories. One type of technique aims to improve multiple users' models performance in general, whereas the other type aims to improve individual user model performance.

\subsection{Multi-user Adaptation}
One of the most popular and noticeable approaches to sensor-based HAR multi-user adaptation is federated learning. Federated learning benefits the multiple subjects with data heterogeneity via integrating the subjects' local models into a centre model and sharing the centre model with each subject for local model performance improvement. This is an iterative process, and eventually, the knowledge from client models is shared through model parameters across different clients. 

\subsubsection{Federated Learning}

In the sensor-based HAR multi-user data heterogeneity scenario, federated learning focuses on learning invariant features cross multiple users and the invariant features can also be used in activity classification efficiently. The essence of federated learning is knowledge sharing from multiple users.

Some works focus on the federated learning strategy to deal with the multi-user data heterogeneity problem. Sannara et al.\cite{ek2020evaluation} evaluated and compared the performance of three types of federated learning strategies of FedAvg, FedPer and FedMa. FedAvg (Federated Averaging) aims to aggregate the local model updates from multiple clients by computing the average of the model parameters. FedPer (Federated Personalization) is an extension of FedAvg that takes into account the heterogeneity of the client data by weighting the local model updates based on the similarity between the local and global models. FedMa (Federated Moving Average) is another extension of FedAvg that uses a moving average to smooth out the fluctuations in the local model updates and reduce the noise in the global model. The result shows that FedAvg achieves the best generalization performance on clients. Presotto et al. \cite{presotto2022fedclar} proposed a novel federated learning strategy that clusters similar users to form multiple groups to mitigate the side effect of different data distribution of users. The intuition idea is that similar users share similar sensor data patterns, and if two client models share similar weights, then the corresponding users are likely to have similar patterns of activities and a small data heterogeneity degree. Therefore, the last several layers’ weights of client models are used for calculating the cross-user similarity score.

Chen et al. \cite{chen2020fedhealth} proposed a CNN-based federated learning framework with an extra data distribution component to mitigate the data heterogeneity issue. The general framework is the same as the common federated learning framework. Firstly, the centre dataset is used to train the centre model. Then, the centre model is distributed to all the clients and each of them can train their individual models on their own labelled data. After that, the client models are pushed to the cloud to iteratively update the centre model. Lastly, each client re-trains its customized model by combining the newly trained cloud model. To further reduce the divergence between the centre model and client models, a correlation alignment layer is added before the softmax layer to align the second-order statistics between the centre model and client models.

\subsubsection{Multi-task Learning}
Multi-task learning can be applied to solve multi-user adaptation and to mitigate the data heterogeneity issue across multiple users. Because focusing only on a single model may ignore the potential information in some related tasks that may promote the target task. By sharing parameters between different tasks to a certain extent, the original task may be better generalized to benefit all users.

Chen et al. \cite{chen2019distributionally} presented a multi-task learning method to solve multi-user data heterogeneity and improve the overall performance for all users simultaneously. This method aims to reduce person-specific discrepancy by aligning the marginal distributions of the labelled data and the unlabelled data and preserving task-specific consistency by generating paired data and maintaining consistency across their features. To achieve the goal, there are four tasks for learning synchronously: (1) the user adversarial loss that forces a reduction in the distribution divergence of the latent features of labelled and unlabelled data via Jensen-Shannon divergence; (2) the reconstruction loss that learns two decoders to reconstruct input vectors from latent features via autoencoder; (3) the latent consistency loss is a constraint that avoids losing the task-specific information during training; (4) the final prediction loss that encourages the encoder to learn discriminative features and ensures a powerful label predictor is trained.

\subsubsection{Subjects Clustering}
Subjects clustering highlights sharing data within similar subjects to form clusters for reducing the sensor-based HAR data heterogeneity among all the users. Therefore, the shared model in each cluster has a good generalization and can be applied to the user without labelled data in this cluster. 
Sztyler et al.\cite{sztyler2017position} researched four cross-subjects approaches, namely  Randomly, LOSO (Leave-One-Subject-Out), Top-Pairs, and Physical. Randomly means the data are chosen at random except the target user from subjects for training the classifier. LOSO strategy is used to build a classifier for each subject that relies on all available labelled data except the target person. Top-Pairs means comparing the subjects pairwise to find the best match subject for each subject. Only in this case, new user labelled data is a requirement. In the Physical approach, the choice is made based on the idea that people with the same fitness level should have similar patterns and data distribution divergence should be small. So only similar fitness level users are clustered together. Then, all the labelled data in the same cluster are trained together to improve the general model performance and mitigate the data heterogeneity issue even if some users have no labelled data. Similar to the above fitness level clustering method, Ferrari et al. \cite{ferrari2020personalization} explored the relationship between physical characteristics and signal similarity across different users. Physical characteristics similarity such as age, weight and height, sensor signal similarity and the combination of these two are under consideration.

\subsubsection{Batch Normalization}
Batch normalization is a deep learning trick. It attracts more and more attention for solving sensor-based HAR data heterogeneity issue because of the effectiveness in reducing covariate shift within deep neural network layers. Mazankiewicz et al.\cite{mazankiewicz2020incremental} added a Domain Alignment Batch Normalization layers to align the feature distributions of all subjects no matter whether they are from source domains or target domains in hidden neural network layers for reducing the multi-user data heterogeneity negative effect.  Domain Alignment Batch Normalization reduces the covariate shift across all the users. Therefore, the deep neural network learns a feature transformation that maximizes class separability while making the domains overlap. 

\subsubsection{Methods Comparison}

\begin{table}[h!] \scriptsize
\caption{Multi-user Adaptation Methods Comparison.\label{Multiple_Users_Adaptation_Comparison_table}}
\begin{tabularx}{\textwidth}{CCCCCC}
\toprule 
\textbf{Technique} & \textbf{Pros} & \textbf{Cons}  & \textbf{Activity Categories}\\ \hline
Federated Learning & good users information sharing mechanism & generalized model may not work well for a specific user & daily living activity, sports fitness activity, atomic activity, composite activity\\ \hline
Multi-task Learning & good model generalization capability & extra efforts for elaborated multiple tasks & atomic activity, composite activity\\ \hline
Subjects Clustering & no need for labelled data & inappropriate selection of clustering similarity metric may drop the performance & daily living activity, composite activity\\ \hline
Batch Normalization & no extra computation cost & not suitable for conditional distribution alignment & daily living activity, sports fitness activity\\
\bottomrule
\end{tabularx}
\end{table}

As shown in Table ~\ref{Multiple_Users_Adaptation_Comparison_table}, federated learning is originally designed as distributed machine learning for shared data security and privacy. It is also a good user information sharing mechanism that can be applied to solving subject data heterogeneity issue. However, the generalized model may not work well for some specific users if the users' data distribution is very different from the general model after the inappropriate selection of a federated learning strategy. Multi-task learning highlights the model generalization capability to solve multi-user data heterogeneity issue via sharing parameters between different tasks for the original task to be better generalized to benefit all the users. However, how to elaborate the multiple tasks for all the users' performance improvement still need extra effort. Subjects clustering follows the unsupervised learning approach with no requirement for labelled data. The key element is the appropriate selection of the clustering similarity metric which is highly related to the task of reducing data heterogeneity cross users. Otherwise, inappropriate clustering metrics may cause a performance drop for all the users. Batch normalization is originally designed as a deep neural network training trick. It is used to reduce covariate shift within deep neural network layers and thus to stabilize and accelerate training. Therefore, the usage of batch normalization for handling data heterogeneity issue has no extra computation cost. However, this method is a coarse-grained alignment for the general data adaptation without considering the conditional distribution alignment. In other words, if two users' classifiers' decision boundaries are different, batch normalization is hard to work well. The corresponding human activity categories in each approach are also listed in Table ~\ref{Multiple_Users_Adaptation_Comparison_table}. Federated learning covers all four types of activity categories. In contrast, multi-task learning is more commonly applied to handle atomic activity or composite activity. Subjects clustering focuses on daily living activity and composite activity. Batch normalization considers daily living activity and sports fitness activity.

\subsection{Single User Adaptation}
Single user adaptation often refers to personalized model learning. Personalisation means the process of capturing a user's personal characteristics \cite{abdallah2018activity}. The dominant learning paradigm in this category is transfer learning, which transfers knowledge and aligns feature distribution from the source user(s) to the target user. In this way, sensor-based data heterogeneity between source user(s) and target user can be mitigated that benefits the model performance of the target user. Here, we divided transfer learning into more detailed categories, namely, samples clustering, subspace mapping, deep neural network and ensemble learning.

\subsubsection{Samples Clustering}
Some works focus on samples clustering to assign the labels corresponding to the source user(s)' samples to their neighbouring samples from the target user. Fallahzadeh and Ghasemzadeh \cite{fallahzadeh2017personalization} presented to train a general model first from the source domains with labelled data. Once a new user emerges, the activity learner module extracts feature-based similarities between the source domain and target domain for labelling unseen data from the user. Finally, the personalized model is trained via the labelled data. Here, clustering and weight bipartite matching are applied to unsupervised data annotation. Data are clustered first, and then the relationship between clusters and labels is learned by weight bipartite for label propagation. 

Vo et al. \cite{vo2013personalization} proposed a method to transfer a trained model from one user to another user via personalized adjustment. First, the labelled samples of person A are used to train a SVM classifier. Second, the model is directly used to classify the unlabelled samples of person B to get the data annotations. The unlabelled samples of person B are also clustered as the same number of the activity classes by K-means and K-medoids algorithms. Third, only the confident samples are selected from the clustering, and the confident samples’ labels are the same as the cluster centre points’ labels. The cluster centre points’ labels are attained by model classification in the second step. 

Zhao et al.\cite{zhao2018user} used a similar framework as \cite{vo2013personalization} but considered the drawback of the K-means algorithm, where the clustering results and the number of iterations are dependent on the initial cluster centres, and the iterative process can be very slow to converge with a bad initialization. So, they added an extra restricted condition of a pre-collected labelled dataset with great similarity to the new user’s unclustered and unlabelled dataset. Then the initial centre points with labels are set to be the cluster centres of this dataset before conducting K-Means clustering steps on the unclustered dataset. After the clustering step, each sample is automatically labelled using the same annotation as the labelled initial point in each cluster. Different from common confident samples selection that ignores the effect of relative density on samples selection, they used local outlier factor technique for removing some outliers with low relative density in each cluster. In the model classification training phase, multivariate Gaussian distribution classifier is used because of its low computational complexity for wearable device. 

\subsubsection{Subspace Mapping}
Subspace Mapping is based on the assumption that there is a common subspace among different users and the features in this subspace are invariant cross users. Therefore, finding the subspace and mapping different domains to the subspace can handle the subject data heterogeneity issue.

Saeedi et al. \cite{saeedi2018personalized} proposed a manifold-based subspace mapping method for single user adaptation. This method links source and target subjects by constructing manifolds from feature subspace representation of the source subject(s) via locally linear embedding algorithm. The algorithm assigns labels to the unlabelled data in the target subject using the manifolds learned from the source subject(s) by Hausdorff metric for measuring the closeness of a manifold to the mapped data. The newly labelled data is used to train a personalized HAR model by RF classifier. 

Fu et al. \cite{fu2021personalized} proposed a new transfer learning algorithm that combines improved pseudo-labels and the joint probability domain adaptive method. The improved pseudo-labels method uses supervised locality preserving projection to learn the projection matrix for mapping the source domain and target domain to the same subspace. Only the source domain is used to obtain the projection matrix at the beginning and then assign pseudo labels to the target domain. The nearest class prototype and structured prediction are applied to the label target domain. Different from common methods such as transfer component analysis and joint distribution adaptation that consider only either the marginal probabilities or conditional probabilities, joint probability domain adaptive derives from the inequality assumption of joint probabilities, which means improved performance on the between-domain transferability and the between-class discrimination. 

Liu et al. \cite{liu2022long} focused on extracting constant information cross users. For example, the duration of a healthy adult's every single motion, such as jumping, sitting down, standing up, jogging (one gait), turning right, etc., clearly falls within a time length of about 1 to 2 seconds and is normally distributed among individuals. With the relatively constant features cross users, data heterogeneity issue can be mitigated. This paper analyzed the duration statistics and distribution of 22 fundamental individual movements of both everyday activities and sports activities. Moreover, in another work \cite{liu2021motion}, human activity is partitioned into a sequence of distinguishing states for model generalization and finding relative invariant information cross users.

\subsubsection{Deep Neural Network}

The deep neural network has the capability of non-linear mapping and feature extraction. More and more works focus on exploring deep neural network transfer learning for user adaptation. Ding et al. \cite{ding2018empirical} did an empirical study on CNN-based deep transfer learning between users. Multiple-kernel Maximum Mean Discrepancy, Domain-adversarial Neural Network and Wasserstein Distance are three common CNN-based transfer learning components which are used to reduce the sensor-based HAR data heterogeneity degree cross domains that are applied in this paper. Moreover, centre loss integrated into Multiple-kernel Maximum Mean Discrepancy loss is utilized to improve the cohesion of inner-class feature distribution among the source domain and target domain to further improve the target user model adaptation performance. 

Some researchers loose the restriction of no labelled data at all from the target domain, which needs a few labelled data that can cover all the activity classes from the new user. Fine-tuning method is the popular method in deep neural network transfer learning when labelled target domain data is available. This method assumes the lower layers of the neural network can extract common features among all users and the sensor-based HAR data heterogeneity only occurs on the higher layers which is specific for each user. Rokni et al. \cite{rokni2018personalized} introduced a CNN deep learning model for cross-subject transfer learning fine-tuning. First, a deep network is trained by a group of users as the source domains. Then, the lower layers of the network are fixed, and the upper layers are re-trained with a few labelled data from the target user.

\subsubsection{Ensemble Learning}
For user adaptation, ensemble learning provides a flexible adaptation approach for solving subject data heterogeneity. It can be used to train a weak classifier for the new user, then a weight update is implemented to borrow the previous knowledge from other weak classifiers. Casale et al. \cite{casale2012personalization} trained a general classifier using an ensemble method of AdaBoost with a large amount of data from many subjects. When a new user appears, the general classifier is applied to adjust its weight and adds the new weak classifier to its weak classifier set with the labelled data from the new user. 

Hong et al. \cite{hong2015toward} trained a Bayesian network, a naïve Bayes classifier and a SVM separately for each user’s activities to construct a pool of activity models. Each model is trained as a binary classifier with the labelled data. When a new user appears, the fitness of all models in the pool is measured with the new user’s small amount of labelled data. The models in the pool with high fitness values are selected to build a hybrid model for the new user. High fitness means the small data heterogeneity between the selected model and the target user, and the fitness value is the weight of the hybrid model that shows how much of other users' knowledge needs to be transferred to the new user.

\subsubsection{Methods Comparison}

\begin{table}[h!] \scriptsize
\caption{Single User Adaptation Methods Comparison.\label{Single_User_Adaptation_Comparison_table}}
\begin{tabularx}{\textwidth}{CCCCCC}
\toprule 
\textbf{Technique} & \textbf{Pros} & \textbf{Cons} & \textbf{Activity Categories}\\ \hline
Samples Clustering & no need for labelled target user data & highly depend on the samples clustering similarity metric & daily living activity, sports fitness activity, composite activity\\ \hline
Subspace Mapping & no need for labelled target user data & optimal subspace dimension is hard to decide & daily living activity, sports fitness activity\\ \hline
Deep Neural Network & strong common feature extraction capability cross users & computationally expensive & daily living activity, sports fitness activity\\ \hline
Ensemble Learning & flexible adaptation with updated weight & need to train a new weak classifier every time when a new user appears & daily living activity, sports fitness activity\\
\bottomrule
\end{tabularx}
\end{table}

As shown in Table ~\ref{Single_User_Adaptation_Comparison_table}, both samples clustering and subspace mapping have no requirement of labelled target user data. However, samples clustering has the drawback of its high dependency on the samples clustering similarity metric. For the subspace mapping, there is no ideal technique to select the optimal subspace dimension which is important to achieve a good performance for target user adaptation. Deep transfer learning method has a strong common feature extraction capability cross
users considering the structure of deep neural network. However, the computation cost is expensive and model training needs more training tricks. Ensemble learning is a flexible adaptation approach for solving subject data heterogeneity. However, it needs to train an extra weak classifier every time when a new user appears. The corresponding human activity categories in each approach are also listed in Table ~\ref{Single_User_Adaptation_Comparison_table}. Samples clustering, subspace mapping, deep neural network and ensemble learning all focus on daily living activity and sports fitness activity. In addition, samples clustering can also be applied to the composite activity.
\section{Spatial Data Heterogeneity}

Sensor-based HAR heterogeneity in spatial dimension includes two categories: environmental heterogeneity and body position heterogeneity. Environmental heterogeneity is associated with the fact that different environments have different sensor layouts. For example, we have the task of identifying the householders' cooking activities in two smart houses with different sensor devices and/or sensor deployment positions. Environmental heterogeneity contains two types of data sources: ambient sensor and device-free. The ambient sensor is the traditional and most common data source that needs to be installed in advance for sensor reading and HAR. Device-free is a new research direction that focuses on analysing wireless electromagnetic wave signal patterns for sensor-based HAR. In contrast, body position heterogeneity refers to cases where sensor data patterns can be different when the same sensor is deployed on different positions of a human body, such as arm and leg. Different data patterns can happen even when people perform the same activity. For instance, the same sensor embedded in a shoe and in a smartwatch has different data patterns when the user runs.

\subsection{Environmental Heterogeneity}

Many works were mainly focused on cross-subjects in the same environment. The research on the cross-subjects and cross-sensors in different environments is less explored. Sensor-based HAR data heterogeneity in environment layout refers to the different number of sensors, sensor deployment layouts and other variations due to changing physical environments. 

\subsubsection{Ambient Sensors}

Smart homes are typically deployed with ambient sensors to identify human activities and describe environmental features. The common method is to find a way to map the source sensor network to the target sensor network based on the similarity. Essentially, the approaches addressing this issue typically discover the common latent characteristics and the feature relationship among the different physical environments. Therefore, the mainstream method is feature transformation under the transfer learning paradigm \cite{zhuang2020comprehensive} across homes in solving the sensor-based HAR spatial data heterogeneity caused by different physical environments. Most of the methods focus on feature extraction in temporal, spatial and object interaction perspectives.

\paragraph{Transfer Learning}

Chen et al. \cite{chen2017activity} combined four machine learning techniques for common features discovery and feature transformation. PCA is used in the source house and target house separately to unify the heterogeneous features to the same dimensions for reducing the data heterogeneity. Then, Jensen-Shannon divergence (JSD) is applied to measure the feature similarity between the source house and the target house. Lastly, Gale-Shapley (GS) algorithm matches features based on the above similarity value between the source house and the target house for the feature transformation.

Feuz and Cook \cite{feuz2015transfer} presented a method called feature-space remapping (FSR) to solve spatial data heterogeneity issue. The FSR algorithm focuses on the challenge of the source and target domain coming from different feature spaces that is very common in spatial data heterogeneity considering the different number of sensors or different sensor layouts deployed in different houses. Compared to traditional transfer learning which is transferring from source to target, or transferring source and target domain to a common feature space, FSR learns a mapping from each dimension in the target feature space to a corresponding dimension in the source feature space. FSR firstly selects and computes meta-features such as the average sensor event frequency over one hour, and the mean and standard deviation of the time between sensor events. Then, a many-to-one similarity mapping matrix is generated between the source house and the target house by calculating each feature-feature pair based on meta-features. Except for the basic FSR version, genetic algorithm FSR and greedy search FSR \cite{feuz2014heterogeneous} were presented for further reducing the data heterogeneity.

The above transfer learning methods are based on the data-driven approach. There are also some works that focus on the knowledge-driven approach, especially ontology-based methods which abstract activities as models of multiple entities, attributes, components and relationships \cite{rodriguez2014survey}. Because manual knowledge engineering conceptualises and abstracts activity recognition semantic and context process, the knowledge-based method naturally has the capacity for generalization and dealing with spatial data heterogeneity. For example, in the heterogeneous deployment of sensors in homes, despite different physical layouts, the functional areas can be abstracted as the bedroom, kitchen, and toilet, which share similar characteristics.  

Ye et al. \cite{ye2014usmart} proposed a knowledge-driven method by introducing semantic relations in object sensor events. The ontological model includes four parts: objects, location, sensor and activity ontology. Objects and location ontology are domain-independent because they typically occur in every household. For example, every house has a sleeping area and tableware. Sensor ontology and activity ontology depend on the specific application scenario—for instance, different sensor deployments based on environment and different activities of interest. Next, the semantic similarity between sensor events is measured by temporal, spatial and object features from the ontology model. After extracting all the unique sequences and mapping each activity, K-means clustering is applied to each activity sequence to discover the representative pattern of the activity based on semantic similarity. Last, sequential patterns are used in a new environment for activity recognition. 

\paragraph{Multi-view Learning}

Unlike the above works under the transfer learning paradigm, which has a transformation direction from the source to the target domain, Ye \cite{ye2018slearn} extended to multi-view learning to handle environment layout heterogeneity synchronously across multiple houses. Here, each house can be treated as a view and each view provides a part of information. The assumptions are that each house can only have a small fraction of data being labelled, and different houses sensors' data correlate in feature space. For each house, a classifier is trained and an uncertainty measure is estimated using a small amount of labelled data. A set of unlabelled examples are generated by randomly selecting from all the houses' unlabelled data. Each unlabelled example from the set is sent to its corresponding classifier to get its class probabilities. In addition, a confidence value is given to the unlabelled example via the uncertainty measure. Then, the most confident examples from the set with their predicted labels are used to complement the limited labelled data. A final classifier is trained based on the new and the limited labelled data for all houses. In this way, spatial data heterogeneity is solved not in a house-house pair way, but by learning a general model to cover the complete data distribution of all the houses' sensors data.

\paragraph{Methods Comparison}

\begin{table}[h!] \scriptsize
\caption{Ambient Sensors Environmental Heterogeneity Methods Comparison.\label{Ambient_Sensors_Environmental_Heterogeneity_table}}
\begin{tabularx}{\textwidth}{CCCCCC}
\toprule 
\textbf{Technique} & \textbf{Learning Paradigm} & \textbf{Learning Approach} & \textbf{Pros} & \textbf{Cons} & \textbf{Activity Categories}\\ \hline
PCA + GS + JSD\cite{chen2017activity} & Transfer Learning & Data-driven & less computation time & one-on-one features mapping may cause information loss & composite activity\\ \hline
FSR\cite{feuz2015transfer}\cite{feuz2014heterogeneous} & Transfer Learning & Data-driven & can relate features in heterogenous feature-spaces & highly depend on the suitable choice of meta-features & composite activity\\ \hline
Ontology-based Model\cite{ye2014usmart} & Transfer Learning & Knowledge-driven & good model generalization & location ontology may be inefficient if different house types & composite activity, daily living activity\\ \hline
Sharing Model\cite{ye2018slearn} & Multi-view Learning & Data-driven & solving data heterogeneity cross all houses together & not work if different domains have a different set of activities & composite activity, daily living activity\\
\bottomrule
\end{tabularx}
\end{table}

As shown in Table ~\ref{Ambient_Sensors_Environmental_Heterogeneity_table}, PCA + GS + JSD method has less computation time because of the low computation complexity of this method. However, the feature mapping mechanism follows the one-on-one pairs way which may cause the information loss issue. For instance, sensor A is deployed in the bedroom of a small house and the bedroom is also used as an office area. Sensor B is deployed in the bedroom and sensor C is deployed in the office room in a big house. In this case, sensor A's data features should follow the one-to-many mapping way that maps sensor B and sensor C. 

FSR is a robust method that can solve environmental heterogeneity and feature mapping even if the heterogeneous feature spaces are in different houses, which is an ideal method considering the different sensor layouts in different houses. However, FSR method highly depends on the choice of meta-feature and may cause target house model performance to drop dramatically in some scenarios. For instance, people have different daily routine habits; someone prefers to sleep early while someone tends to sleep late. In this case, the meta-feature of average sensor event frequency over a certain time period may not be representative across different homes. 

Ontology-based model is a knowledge-driven method, abstracts environments as models of objects, location, sensor and activity ontology. Objects and location ontology are domain-independent because they typically occur in every household. In this way, environmental heterogeneity can be solved based on the definition of the same functional area. However, the model generation of location ontology may be inefficient if different house types. For example, the sleeping area and leisure area may overlap in a crowded urban house compared to a country house, which leads to an inappropriate design of location ontology.

Different from the above transfer learning paradigm, sharing model follows the paradigm of multi-view learning that highlights solving data
heterogeneity cross all houses
at the same time. This method combines the information from all the houses together and benefits all the houses as well. However, the success of this method is based on the assumption that different domains have the same set of activities. If every house have no overlapped activities, there is no useful information to share across the houses.

The corresponding human activity categories in the above four approaches are listed in Table ~\ref{Ambient_Sensors_Environmental_Heterogeneity_table} as well. All four types of techniques have been applied to composite activity, with exceptions on ontology-based and sharing model approaches which can also be applied to the daily living activity.

\subsubsection{Device-free}

In recent years, device-free activity recognition has been an emerging research area that focuses on analysing wireless electromagnetic wave signal patterns to classify different activities. Due to the physical phenomenon of electromagnetic reflection, diffraction and refraction, a person’s physical activities lead to changes in the surrounding wireless signal propagation. However, the wireless signal is sensitive and easily disturbed by the change of environments such as different house layouts or signal interference due to the operation of household appliances. Moreover, the research on device-free based HAR data heterogeneity in environment layout is at an early stage. The dominant technique in this field is deep learning related transfer learning which has the capabilities of automatic feature extraction, denoising and complex model construction.

A deep transfer learning method \cite{jiang2018towards} is applied to extract the subject-independent and environment-independent features for solving device-free environmental data heterogeneity issue. To eliminate the interference of specific environment and subject noise, a CNN-based deep domain discriminator is used to force the feature extractor to remove domain-dependent activity features. In addition, confidence control constraint and smoothing constraint are proposed to handle overfitting and fluctuating latent space issue, and balance constraint confines the percentage of each activity in the final prediction to be the same as the percentage of prior knowledge. The highlight of the research is the evaluation and comparison of numerous device-free data types, including ultrasound, mmWave, visible light and Wi-Fi signals. 

With a similar idea of finding domain-independent features, Zheng et al.\cite{zheng2019zero} proposed a deep transfer learning method and constructed a cross-domain gesture recognition feature called Body-coordinate Velocity Profile (BVP) under the assumption of different locations and orientations relative to Wi-Fi links and environments. Time-frequency analysis and motion tracking originating from channel state information (CSI) signal are used to generate Doppler Frequency Shift profile, orientation and location information. Then, the BVP feature is refined from this information by compressed sensing BVP estimation. Last, a GRU + CNN deep learning one-fits-all model is applied to extract temporal and spatial features for gesture recognition. In addition, the dataset of Wi-Fi gesture recognition called Widar3.0 is released to the research community. 

Wang et al. \cite{wang2020cross} developed a CNN-based deep transfer learning model to transfer target domain feature distribution to the source domain distribution. The ingenious design of the method is the process of maximizing the decision discrepancy of the classifier for tightly encircling the feature distribution by classifier model adjustment. An adversarial technique is developed to guide the feature extractor which is a CNN-based deep neural network to move the target features to the distribution of the source features. Then, an alignment technique is applied to relocate the target features to the distribution centre of the source domain for further attaining a uniform distribution of the target domain.

Compared to the above methods using spatial and temporal feature extraction directly without particular intention, Shi et al. \cite{shi2020environment} aim to enhance the activity-dependent feature and ignore the activity-unrelated feature such as signal reflection from the background static objects. 
The raw CSI signal contains multiple channel paths from static background objects and hence a lot of activity-unrelated information. Such information is generally environment-dependent and the spatial data heterogeneity issue is obvious. To solve the data heterogeneity issue, they used a linear recursive operation to construct the CSI signal for static objects and then subtract it from the received signal to get activity-unrelated information. Then, they extracted useful features from the filtered CSI, including two parts of activity-related information feature and temporal correlation feature for the following activity recognition task. 

\begin{table}[h!] \scriptsize
\caption{Device-free Environmental Heterogeneity Methods Comparison.\label{Device_Free_Environmental_Heterogeneity_table}}
\begin{tabularx}{\textwidth}{CCCCCC}
\toprule 
\textbf{Deep Transfer Learning Technique} & \textbf{Labelled New Data} & \textbf{Pros} & \textbf{Cons} & \textbf{Activity Categories}\\ \hline
Domain Discriminator\cite{jiang2018towards} & No & good at marginal distribution alignment & coarse-grained alignment without considering the conditional distribution alignment & daily living activity, composite activity\\ \hline
BVP\cite{zheng2019zero} & No & the extracted features are domain independent & uncertainty of the number of wireless links are required to uniquely recover the BVP & atomic activity\\ \hline
Adversarial Method\cite{wang2020cross} & No & no need for labelled samples in the target house & may cause overfitting after the step of maximizing the classifier's decision discrepancy & atomic activity\\ \hline
Activity-unrelated Information Filter\cite{shi2020environment} & Yes & good generalization regardless of the environments change & need to generate a CSI signal pattern for each new environment & atomic activity, composite activity, daily living activity\\
\bottomrule
\end{tabularx}
\end{table}

As shown in Table ~\ref{Device_Free_Environmental_Heterogeneity_table}, the domain discriminator is efficient if two houses have overall data heterogeneity issue and the decision boundaries of the classifiers should be similar to each other, which means it works well in marginal distribution alignment. However, the domain discriminator is only applied at the domain level which is coarse-grained alignment without considering the conditional distribution alignment. Therefore, if the decision boundaries of the classifiers are very different, this method will fail to work. BVP is a newly proposed feature presentation that is domain independent cross houses. Therefore, with this common feature, the environmental heterogeneity issue can be solved. However, the derivation of this feature depends on some signal processing techniques. The uncertainty of the number of wireless links required to uniquely recover the BVP in different environments may cause the computation complexity to be hard to estimate in advance. The adversarial method does not require labelled samples in the target domain, which has a wider application scenario. However, the step of maximizing the classifier’s decision discrepancy may lead to the model overfitting problem and affect the transfer learning performance. Activity-unrelated Information Filter method has a good generalization capability regardless of the environmental change. However, it needs an extra step to generate a CSI signal pattern for each new environment. The corresponding human activity categories in each approach are also listed in Table ~\ref{Device_Free_Environmental_Heterogeneity_table}. BVP and adversarial methods focus on atomic activity. Domain discriminator considers daily living activity and composite activity. Activity-unrelated information filter aims to handle atomic activity, composite activity and daily living activity.

\subsection{Body Position Heterogeneity}

Wearable devices or smartphones activity recognition commonly assume the fixed sensor position on the body to simplify the operation. However, this assumption is sometimes difficult to hold in practical applications because the physiological structure of the human body leads to the very different distribution of sensor data from various body parts even with the same activity. For example, when people are running, the acceleration signal fluctuates intensely in the ankle compared to the waist \cite{wang2017kernel}. There are usually two main methods in this area: 1) position-aware methods that take into consideration the sensor’s position, and 2) position-independent methods that make use of some on-body position-independent features.

\subsubsection{Position-aware Method}

Compared to the sensor-based HAR data heterogeneity issue in different subjects, sensor-based HAR data heterogeneity in body position issue is relatively more manageable because of the limited human positions such as waist, ankle, arm and leg. Enumeration method with standard supervised learning is possible in this field, which means the test dataset belongs to one of the training datasets. Transfer learning is the other learning paradigm that transfers knowledge from one or more body position(s) to the target body position.

\paragraph{Enumeration Method}

Yang et al. \cite{yang2016pacp} also identified the smartphone sensor position first but used the decision tree model. After adjusting the accelerometer data corresponding to the position, SVM is trained for activity classification. To eliminate the influence of smartphone orientation, the coordinate system is converted from tri-axial acceleration directions to vertical and horizontal directions.

Sztyler et al. \cite{sztyler2016body} considered the influence of the different on-body positions of the mobile device and predicted the sensor position based on acceleration data via a random forest classifier. The stratified activity recognition method includes dynamic behaviour (climbing, jumping, running, walking) and static behaviour (standing, sitting, lying) with the features of time (such as entropy, correlation coefficient and kurtosis) and frequency (such as Fourier transformation).

\paragraph{Transfer Learning}

Transfer learning is another learning approach in this area. Wang et al. \cite{wang2018stratified} presented a transfer learning method for position-aware body position heterogeneity issue. The idea of the framework includes three steps of majority voting for generating pseudo labels for the target body position via the classifiers trained on source body position, intra-class transfer into the same subspace between source body position and target body position via maximum mean discrepancy similarity calculation, and second annotation in the target domain. The second and third steps are the iterative refinement of the Expectation-Maximization (EM) algorithm until convergence. As an extension work of \cite{wang2018stratified}, Chen et al. \cite{chen2019cross} further add a Stratified Domain Selection component that can select the most similar source body position to the target body position. It is a greedy algorithm that exploits the local properties of the multiple source body positions via the Radial Basis Function (RBF) kernel method. 

Similar to Chen et al.’s algorithm of choosing the right source domain for better performance and preventing the negative transfer, Wang et al. \cite{wang2018deep} made a source body position selection with a different distance measurement method called context activity distance. It is an overall distance including kinetic distance for approximating the signal relationship between source body positions and target body position and semantic distance for indicating the spatial relationship between the source domains and target domain. Deep learning is applied to transfer knowledge and activity recognition, and Maximum Mean Discrepancy is the last step used to reduce the discrepancy between domains for the adaptation layer. 

Rokni et al. \cite{rokni2018autonomous} proposed a transfer learning method to meet the requirement of model adjustment in real-time when new sensor deployment on body position without labelled data. The existing sensors are treated as source domains, and the emerging sensors are treated as the target domain. Source domain data is firstly utilized to generate a likelihood function of existing sensors' activity classification via calculating the occurrence rate of each pair of the predicted label and actual label at a certain time point. Simultaneously, target data use this likelihood function as its classifier and combines the source data and its own data to form a new dataset. Clustering is implemented on this new dataset and the number of cluster groups is the same as the number of activity classes. Afterwards, a mapping function is learned to link the weight relationship between the clusters and activity classes. Finally, the best matched class is assigned to the corresponding cluster and the instances in it for the autonomous annotation. 

\paragraph{Methods Comparison}

\begin{table}[h!] \scriptsize
\caption{Position-aware Body Position Heterogeneity Methods Comparison.\label{Position-aware_Body_Position_Heterogeneity_table}}
\begin{tabularx}{\textwidth}{CCCCCC}
\toprule 
\textbf{Technique} & \textbf{Pros} & \textbf{Cons} & \textbf{Activity Categories}\\ \hline
Enumeration Method & high accuracy to identify the body position & require labelled data from all body positions & daily living activity, sports fitness activity\\ \hline
Transfer Learning & no need for labelled target body position data & performance depend on the cross-domain similarity metric & daily living activity, sports fitness activity, composite activity\\
\bottomrule
\end{tabularx}
\end{table}

As shown in Table ~\ref{Position-aware_Body_Position_Heterogeneity_table}, the enumeration method has high accuracy in identifying the
body position. This method does not focus on reducing body position heterogeneity but builds a model for each body position separately. However, the cost is the requirement of the labelled data from all body positions that are rare in the real world. Transfer learning has no requirement for labelled target body position data. However, the success of solving body position heterogeneity depends on the cross-domain similarity metric which is suitable for reducing body position heterogeneity scenario. The corresponding human activity categories in each approach are also listed in Table ~\ref{Position-aware_Body_Position_Heterogeneity_table}. Both the enumeration method and transfer learning focus on daily living activity and sports fitness activity. In addition, transfer learning can also be applied to address composite activity.

\subsubsection{Position-independent Method}

The dominant method here is the multi-view learning that finds some insensitive features via mixed data from all available body positions for attaining model generalization capacity. Zhou et al. \cite{zhou2020phone} proposed a hierarchical method that distinguishes vertical direction coarse-grained activities in the first layer and then identifies fine-grained activities in the second layer. The barometer sensor is used to identify altitude changes for the classification of horizontal walking, upstairs and downstairs. After feature extraction through wavelet transform and Singular Value Decomposition, machine learning models identify six types of upstairs and downstairs with different movement velocities. The models then recognise five fine-grained walking modes and generate domain-independent features without considering the phone position. Almaslukh et al. \cite{almaslukh2018robust} also combined the data from all positions to train a general model. In addition, they did a comprehensive research on sensor-based HAR data heterogeneity in body position.

Unlike the above method treating each body position as a view for the multi-view learning, Wang et al. \cite{wang2017kernel} treated each sensor as a view for the multi-view learning and applied sensor fusion to generate magnitude series to eliminate the orientation information and then extracted statistical and frequency-domain features. After PCA dimension reduction, the obtained principal components are utilized to train the classification model via a mixed kernel-based Extreme Learning Machine algorithm. Afterwards, the most confident samples combined with initial training data are selected to update the ELM classification model. The renewed model gradually adapts to data in previously unseen locations as the emerging new data from a new sensor location.

\section{General Framework for Multiple Heterogeneities}

Some researchers presented a general framework to handle multiple sensor-based HAR data heterogeneity issues with a sufficient and complete experiment. Except for the four categories of sensor-based HAR data heterogeneity issues that were already been discussed in the previous sections: data modality heterogeneity, streaming data heterogeneity, subject data heterogeneity and spatial data heterogeneity. Some researches focus on solving the sensor-based HAR mixed data heterogeneity scenarios. Here, the transfer learning paradigm is the dominant method.

To better align source domain and target domain features, some work needs labelled source domain data and labelled target domain data. Feng et al. \cite{feng2019few} proposed a general cross-domain deep transfer learning framework combined with parameter-based transfer learning and cross-domain class-wise similarity. In the first step, a stacked LSTM network with a feature extractor and a classifier is trained, and the training data is source domain samples. In the second step, the network parameters of the source feature extractor and classifier are transferred to the target network with the same structure. Especially, the weight of transferred source classifier parameters is dependent on a cross-domain class-wise relevance measure which includes cosine similarity, sparse reconstruction and semantic normalized Google distance. Then, the transferred classifier parameters and the source feature extractor parameters are used to initialise the target network. In the last step, fine-tuning is applied to adjust the feature extractor and classifier in the target network. 

Some work only requires labelled source domain data and unlabelled target domain data. Sanabria et al. \cite{sanabria2021unsupervised} proposed a transfer learning framework from the source domain(s) to the target domain via Bi-GAN. The architecture of Bi-GAN consists of two GANs of a generator and a discriminator on the source and target domain, respectively. In the feature space transformation step, both generators are trained to generate fake samples as close as the real samples in the other domains, and both discriminators are binary classifiers to detect whether an input is generated by their corresponding generators or a real sample from the other domains. In the feature distribution alignment step, the transformed features are shifted to the real target data via Kernel Mean Matching to improve classification accuracy. Then, a classifier on the aligned transformed features is trained, and the features’ corresponding labels are inherited from the source domain. Finally, the classifier is used to label data in the target domain. 

\textbf{Domain generalization} is an emerging area of transfer learning that explores how to acquire knowledge from various related domains. It only assumes that samples from multiple source domains can be accessed and target domain can provide no data at all. Limited work has been done in this field. Erfani et al. \cite{erfani2016robust} proposed the Elliptical Summary Randomisation framework for domain generalization comprising a randomised kernel and elliptical data summarization. The data is projected to a lower-dimensional latent space using a randomised kernel, which reduces domain bias while maintaining the data's functional relationship. The Johnson-Lindenstrauss method provides probabilistic assurances that a dataset's relative distances between data points are preserved when randomly projected to a lower feature space. So, ellipsoidal summaries are employed to replace the samples to increase generalisation by reducing noise and outliers in the projected data. The focal distance between pairs of ellipsoids is then used as a measure of domain dissimilarity. Lu et al. \cite{lu2022local} considered global correlation and local correlation of time series data. A local and global features learning and alignment framework is proposed for generalizable human activity recognition.

\section{Public Datasets}

\subsection{General Datasets}

There are some public datasets that can be applied to evaluate various types of sensor-based HAR data heterogeneity issues. Here, we summarized the datasets that are used in more than or equal to two fields of heterogeneity issues. These eight datasets may attract more attention because of their versatility and capability to conduct different types of evaluation under various scenarios.

\begin{table}[h!] \scriptsize
\caption{General Datasets Data Heterogeneity Types Table.\label{tab3}}
\begin{tabularx}{\textwidth}{CCCCCCCC}
\toprule 
Dataset\& Ref. & Data Modality Heterogeneity & Concept Drift (Streaming Data Heterogeneity) & Concept Evolution (Streaming Data Heterogeneity) & Open-set (Streaming Data Heterogeneity) & Subject Data Heterogeneity & Body Position Heterogeneity (Spatial Data Heterogeneity) & Multiple Heterogeneities\\
\midrule
HHAR\cite{stisen2015smart}                                    & \checkmark (3)         & \checkmark (2)         &                   &          & \checkmark (2)              &               & \checkmark (3)                    \\ 
\midrule
MHEALTH\cite{banos2014mhealthdroid}                                                                     & \checkmark (2)         &               &                   &          & \checkmark (1)              & \checkmark (1)         &                          \\ 
\midrule
OPPT\cite{chavarriaga2013opportunity}                                                                 & \checkmark (1)         & \checkmark (2)         & \checkmark (4)             & \checkmark (1)    & \checkmark (1)              & \checkmark (2)         & \checkmark (6)                    \\ 
\midrule
WISDM\cite{kwapisz2011activity}                                         &               & \checkmark (2)         &                   &          & \checkmark (2)              &               & \checkmark (1)           \\ 
\midrule
UCIHAR\cite{anguita2013public}                                                                     &               & \checkmark (1)         & \checkmark (1)             &          & \checkmark (5)              &               & \checkmark (3)                    \\ 
\midrule
PAMAP2\cite{reiss2012introducing}                                                                      &               &               & \checkmark (5)             &          & \checkmark (4)              & \checkmark (2)         & \checkmark (6)                    \\ 
\midrule
DSADS\cite{altun2010comparative}                                &               &               & \checkmark (1)             &          & \checkmark (5)              & \checkmark (2)         & \checkmark (4)                    \\ 
\midrule
RealWorld\cite{sztyler2016body}                                                                &               &               &                   &          & \checkmark (3)              & \checkmark (2)         &   \\                      
\bottomrule
\end{tabularx}
\end{table}

Table~\ref{tab3} shows the corresponding research and evaluation areas of specific data heterogeneity issues that each dataset can be applied. The number in each bracket means the number of papers that use this dataset in the specific field. The key information of these eight datasets is also summarised in Table~\ref{tab4}.

\begin{table}[h!] \scriptsize
\caption{General Datasets Detailed Information Table.\label{tab4}}
\begin{tabularx}{\textwidth}{m{0.15\textwidth}m{0.15\textwidth}m{0.1\textwidth}m{0.1\textwidth}m{0.1\textwidth}m{0.4\textwidth}}
\toprule 
Dataset \& Ref. & Sensor 

modalities & No. of 

sensors & No. of 

participants & No. of 

Activities & Activity Categories\\
\midrule
HHAR\cite{stisen2015smart}  & Accelerometer, Gyroscope & 36 & 9 & 6 & Daily living activity, 

Sports fitness activity \\ \hline
MHEALTH\cite{banos2014mhealthdroid}  & Accelerometer, Gyroscope, magnetometer, electrocardiogram & 3 & 10 & 12 & Atomic activity, 

Daily living activity, 

Sports fitness activity
\\ \hline
OPPT\cite{chavarriaga2013opportunity} & Acceleration, rate of turn, magnetic field, reed switches & 40 & 4 & 17 & Daily living activity, 

Composite activity\\ \hline
WISDM\cite{kwapisz2011activity}  & accelerometer, gyroscopes & 1 & 33 & 6 & Daily living activity, 

Sports fitness activity\\ \hline
UCIHAR\cite{anguita2013public} & Accelerometer, Gyroscope & 1 & 30 & 6 & Daily living activity\\ \hline
PAMAP2\cite{reiss2012introducing} & Accelerometer, Gyroscope, magnetometer, temperature & 4 & 9 & 18 & Daily living activity, 

Sports fitness activity, 

Composite activity\\ \hline
DSADS\cite{altun2010comparative} & Accelerometer, Gyroscope, magnetometer & 45 & 8 & 19 & Daily living activity, 

Sports fitness activity\\ \hline
RealWorld\cite{sztyler2016body} & Acceleration & 7 & 15 & 8 & Daily living activity, 

Sports fitness activity\\ 
\bottomrule
\end{tabularx}
\end{table}

\subsection{Specific Datasets}
There are some public datasets designed for a specific and particular sensor-based HAR heterogeneity problem. Here, we summarized the specific datasets with more than or equal to two cited papers. These eleven datasets may also be considered when the researchers want to dive into the specific sensor-based HAR data heterogeneity problem.

\begin{table}[h!] \scriptsize
\caption{Specific Datasets Data Heterogeneity Types Table.\label{tab5}}
\begin{tabularx}{\textwidth}{CCCCC}
\toprule 
Data Modality & Concept Evolution & Different Subjects & Environment Layout & Multiple Heterogeneities \\
\midrule
UTD-MHAD\cite{chen2015utd} & Exercise Activity\cite{cheng2013nuactiv} & USC-HAD\cite{zhang2012usc} & van Kasteren\cite{van2011human} & Skoda\cite{zappi2008activity} \\
 & TUD\cite{huynh2008discovery} & Shoaib\cite{shoaib2014fusion} & CASAS\cite{cook2012casas} &  \\
 &  & SHAR\cite{micucci2017unimib} & Widar3.0\cite{zheng2019zero} &  \\
 &  & MobiAct\cite{vavoulas2016mobiact} &  &  \\
 &  & Motion Sense\cite{malekzadeh2018protecting} &  & \\
\bottomrule
\end{tabularx}
\end{table}

Table~\ref{tab5} shows the specific data heterogeneity field and their corresponding problem field-dependent datasets. The key information of these eleven datasets is also summarised in Table~\ref{tab6}.

\begin{table}[h!] \scriptsize
\caption{Specific Datasets  Detailed Information Table.\label{tab6}}
\begin{tabularx}{\textwidth}{m{0.15\textwidth}m{0.15\textwidth}m{0.1\textwidth}m{0.1\textwidth}m{0.1\textwidth}m{0.4\textwidth}}
\toprule 
Dataset \& Ref. & Sensor 

modalities & No. of 

sensors & No. of 

participants & No. of 

Activities & Activity Categories\\ 
\midrule
Exercise 

Activity \cite{cheng2013nuactiv} & Accelerometer, gyroscope & 3 & 20 & 10 & Sports fitness activity\\ \hline
UTD-MHAD \cite{chen2015utd} & Accelerometer, Gyroscope, RGB camera, depth camera & 3 & 8 & 27 & Atomic activity, 

Daily living activity, 

Sports fitness activity, 

Composite activity\\ \hline
Shoaib \cite{shoaib2014fusion} & Accelerometer, Gyroscope & 5 & 10 & 7 & Daily living activity,

Sports fitness activity\\ \hline
TUD \cite{huynh2008discovery} & Accelerometer & 2 & 1 & 34 & Daily living activity, 

Sports fitness activity, 

Composite activity\\ \hline
SHAR \cite{micucci2017unimib} & Accelerometer & 2 & 30 & 17 & Atomic activity, 

Daily living activity, 

Sports fitness activity\\ \hline
USC-HAD \cite{zhang2012usc} & Accelerometer, Gyroscope & 1 & 14 & 12 & Daily living activity, 

Sports fitness activity\\ \hline
MobiAct \cite{vavoulas2016mobiact} & Accelerometer, gyroscope, orientation sensors & 1 & 50 & 13 & Atomic activity, 

Daily living activity\\ \hline
Motion Sense \cite{malekzadeh2018protecting} & Accelerometer, gyroscope & 1 & 24 & 6 & Daily living activity\\ \hline
van Kasteren \cite{van2011human} & switches,

contacts, 

passive 

infrared (PIR) & 14(23)(21) & 1(1)(1) & 10(13)(16) & Daily living activity,

Composite activity\\ \hline
CASAS \cite{cook2012casas} & Temperature, 

Infrared 

motion/light sensor & 52(63) & 1(1) & 7(9) & Daily living activity, 

Composite activity\\ \hline
Skoda \cite{zappi2008activity} & Accelerometer & 19 & 1 & 10 & Daily living activity, 

Composite activity\\ \hline
Widar3.0 \cite{zheng2019zero} & Wi-Fi & 7 & 1 & 6 & Atomic activity\\ 
\bottomrule
\end{tabularx}
\end{table}

\section{Future Directions and Conclusion}

The research work surveyed in this review also identified some future research directions that require further discussion.

\textbf{Cross modality knowledge transformation} in sensor-based HAR is a new and promising research direction. There is still very limited work in this area. Its purpose is to build a bridge and find the relationship between different data modalities for taking advantage of knowledge from the source domain, especially under the scenarios of difficult data collection from the target domain. If the gap between some modalities’ feature space is relatively big and these domains have no direct correlations, it may be difficult to achieve cross-modality knowledge transformation via traditional transfer learning methods. Transitive transfer learning (TTL) \cite{tan2015transitive} is a potential solution inspired by the idea of transmissibility for indirect reasoning and learning. This method has the capability of assisting in connecting concepts and passing knowledge between seemingly unrelated concepts. Normally, the bridge is built to link unrelated concepts by introducing intermediate concepts. TTL is an important extension of traditional transfer learning to enable knowledge transformation across domains with huge differences in data distribution. There are two key points needed to be considered: (1) how to select the appropriate intermediate domain as the bridge between the source domain and target domain, (2) how to transfer knowledge efficiently across the connected domains.

\textbf{Adaptive methods for unseen activities} is another relatively new topic in sensor-based HAR at present and can be applied to a more challenging and practical scenario. It aims to learn a model that can generalize to a target domain with one or several differences but related source domain(s). It only assumes that samples from multiple source domains can be accessed and has no access to the target domain. For HAR, this research topic is suitable for various situations. For example, the activity data are originated from multiple existing users (source domains) but hard to obtain from a new user (target domain). In this case, the methods of domain generalization can be explored to handle this type of challenge. Moreover, extra semantic information can be introduced as external supplementary information. Current researches focus on embedding word vectors such as Word2Vec that build the semantic connection between the activity verbs and object nouns to transfer the classification label space to word vectors space. Other types of external supplementary information may also be explored and introduced to help adaptive methods for unseen activities. For example, the human skeleton and muscle structure research in the medical area can be applied to HAR.

\textbf{Shared public datasets} with high-quality and representative applications are the basis for HAR research and are also of critical importance for the research community. Even though there are datasets released in the past decades. They are still insufficient in several specific areas. For instance, there is no dataset offering long-term data streams that include different levels of concept drift and new emerging classes. Cross modality datasets that simultaneously record, for example, RGB /depth video, key points, IMU data, orientation, Wi-Fi, pressure, RFID, PIR, contact sensor data, etc., are also not available.

Human Activity Recognition (HAR) is an essential part of ubiquitous computing, which aims to identify and analyze human behaviour from multi-dimensional observations and contextual information. Many application scenarios include medical treatment, auxiliary environment life, health, fitness and sports monitoring, rehabilitation, home automation, security monitoring, etc. Despite the tremendous research efforts made in the past few decades, there are still many challenges and emerging aspects of activity identification. Most of the latest generation of HAR methods are performed under the assumption of data homogeneity. Data homogeneity means that the data distribution in all datasets is the same. However, actual human activity sensor data is often heterogeneous. Solving sensor-based HAR data heterogeneity leads to improved performance with lower computational cost and helps to build a robust, adaptable, and custom model with less annotation work. With the recent research development in sensor-based HAR data heterogeneity, this paper analyses the research work and discovers areas that still require further investigation. This review paper categorises different types of data heterogeneity in sensor-based HAR applications. Different machine learning techniques developed for each type of data heterogeneity are analyzed, compared, and summarized with their advantages and disadvantages. The available public datasets are also summarized to facilitate future research in sensor-based HAR data heterogeneity.

\section{References}

\bibliographystyle{ACM-Reference-Format}
\bibliography{ref}

\end{document}